\documentclass[
reprint,
 amsmath,amssymb,
 aps,
prb,
]{revtex4-1}

\usepackage[utf8]{inputenc}
\usepackage{graphicx}
\usepackage{dcolumn}
\usepackage{bm}

\usepackage{xspace}
\newcommand{\eg}{e.\,g.,\xspace}
\usepackage{siunitx}
\DeclareMathOperator*{\argmin}{arg\,min}
\usepackage{physics}
\graphicspath{{bilder/}}

\usepackage{tikz}
\usetikzlibrary{calc,arrows}
\usepackage[english]{babel}

\usepackage{ulem}

\begin{document}

\title{Electronic Correlations in Vanadium Revealed by Electron-Positron Annihilation Measurements}

\author{Josef Andreas Weber$^{a}$, Diana Benea$^{b}$, Wilhelm H. Appelt$^{c,d}$,   Hubert Ceeh$^{a}$, \\ Wolfgang Kreuzpaintner$^{a}$, Michael Leitner$^{a,f}$, Dieter Vollhardt$^{e}$, Christoph Hugenschmidt$^{\ast, a,f}$, Liviu Chioncel$^{d,e}$}

\affiliation{$^{a}$Physik-Department, Technische Universit\"at M\"unchen, James-Franck Stra\ss e, 85748 Garching, Germany}
\affiliation{$^{b}$Faculty of Physics, Babes-Bolyai University, Kog\u alniceanustr 1, 400084 Cluj-Napoca, Romania}
\affiliation{$^{c}$Theoretical Physics II, Institute of Physics, University of
Augsburg, 86135 Augsburg, Germany}
\affiliation{$^{d}$Augsburg Center for Innovative Technologies, University of Augsburg,
86135 Augsburg, Germany}
\affiliation{$^{e}$Theoretical Physics III, Center for Electronic
Correlations and Magnetism, Institute of Physics, University of
Augsburg, 86135 Augsburg, Germany}
\affiliation{$^{f}$Heinz Maier-Leibnitz Zentrum (MLZ), Technische Universit\"at M\"unchen, Lichtenbergstr. 1, 85748 Garching, Germany}


\begin{abstract}
The electronic structure of vanadium measured by Angular Correlation of electron-positron Annihilation Radiation (ACAR) is compared with the predictions of the combined Density Functional and Dynamical Mean-Field Theory (DMFT).
Reconstructing the momentum density from five 2D projections we were able to determine the full Fermi surface and found excellent agreement with the DMFT calculations. 
In particular, we show that the local, dynamic self-energy corrections contribute to the anisotropy of the momentum density and need to be included to explain the experimental results.
\\
\\
$ ^{\ast}$\footnotesize{Corresponding author: hugen@frm2.tum.de}
\end{abstract}

\maketitle

\section{Introduction}
The annihilation process of electron-positron pairs in matter is well understood~\cite{West1973,Karshenboim2005}: As a consequence of momentum and energy conservation electron-positron pairs decay in solids  predominantly into two $\gamma$-quanta (photons). The Two Photon Momentum Density [TPMD or $\rho^{2\gamma}({\bf p})$] of the annihilation radiation carries valuable information about the electron momentum density sampled by the positron.
Early annihilation studies in metals already observed a strong electron-positron attraction superposed on the many-body correlations between the electrons~\cite{be.co.50}.

On the theoretical side, the Density Functional Theory (DFT)~\cite{jo.gu.89,kohn.99,jone.15} can describe many-body effects in the ground state of solids only if the exact electron exchange-correlation potential is included.
Since this is not known most computational schemes based on DFT use either the Local Density Approximation (LDA) or the Generalized Gradient Approximation (GGA). 
For paramagnetic correlated ($3d$ and $4f$) electron systems it is well known that LDA/GGA calculations fail to provide the correct ground state properties.
A quantitative theory for the explanation of the electronic structure and the physical properties of such systems has been consistently developed during the last two decades in the form of a combination of density functional theory and Dynamical-Mean Field Theory (DMFT)~\cite{me.vo.89,ge.ko.96,ko.vo.04,ko.sa.06}, which is generally referred to as LDA+DMFT~\cite{ko.sa.06,held.07}. 
In the LDA+DMFT scheme the LDA provides the ab initio, material dependent input (orbitals and hopping parameters), while the DMFT solves the many-body problem for the local interactions. 
Thereby the LDA+DMFT approach is able to compute, and even predict, properties of correlated materials. 
Recently, the methodology to compute the spin-polarized two-dimensional angular correlation of annihilation radiation (2D-ACAR) in combination with LDA+DMFT was developed~\cite{Ceeh2016}.   
This made it possible, for example, to experimentally pinpoint the strength of the local electronic interaction in Ni.\cite{Ceeh2016}

Vanadium is a $3d$ transition metal element with the electronic configuration of [Ar] $3d^3$ $4s^2$.
The electronic structure of vanadium has been intensively studied by electron-positron 
annihilation~\cite{Wakoh1975,Manuel1983,Singh1984,Singh1985a,Matsumoto1986,Pecora1988,Dugdale1994,Major2004a,Hughes2004,Dugdale2013,Shiotani1975,Manuel1982} and, less so, by the de Haas - van Alphen (dHvA) effect~\cite{Phillips1971,Parker1974} and photo-emission~\cite{Pervan1996,Peric1995}. 
Indeed the characteristic Fermi Surface (FS) leads to a very pronounced signal in 2D-ACAR spectra projected along specific directions.
Quantum oscillation experiments were able to resolve the N-hole pockets, but gave very little information about the so-called ``jungle-gym surface'' and no information at all about the first octahedral hole surface\cite{Phillips1971,Parker1974}.
Photoemission studies yielded only three points of the dispersion relation~\cite{Pervan1996,Peric1995}.
With the exception of the work of Manuel~\cite{Manuel1982}, who reconstructed different FS sheets in vanadium from only two 2D projections, previous positron studies refrained from reconstructing the Fermi surface (although at least in one case a reconstruction of $\rho^{2\gamma}(\bf{p})$ along selected directions and planes was calculated~\cite{Pecora1988}), contenting themselves with comparing measured projections to band theoretical calculations. 
Momentum distributions of vanadium obtained by Compton scattering show significant differences to LDA calculations~\cite{Wakoh1975}, in particular at small momenta. 
Various reasons for these discrepancies, such as electron-electron correlations and thermal effects have been discussed~\cite{Tokii2003}.
Band structure calculations within DFT using LDA/GGA and LDA+$U$ are available~\cite{Wakoh1975,Tokii2003}, but calculations within LDA+DMFT, the current state of the art, have not been reported.
So far several questions
, such as the validity of the LDA, LDA+$U$ schemes for the calculation of $\rho^{2\gamma}({\bf p})$ and the significance of the electron correlations, are still open. 

In this paper, we report the full 3D-electron momentum density reconstruction from measurements of the Angular Correlation of electron-positron Annihilation Radiation (ACAR) in vanadium which allows us to identify specific signatures of the Fermi surface.
In particular, we compare the measured and computed results for i) the 2D-ACAR anisotropy data, ii) the Lock-Crisp-West (LCW) back-folded spectrum in the first Brillouin Zone, and iii) the complete Fermi surface.  Our experimental results and calculations based on LDA, LDA+$U$ and LDA+DMFT show that there are clear discrepancies between both LDA and LDA+$U$ with the experiment, and that those discrepancies can be partly resolved by including electronic correlations through the dynamical self-energy of DMFT.

\section{Experimental techniques}
\label{sec:exp}

The 2D-ACAR technique is a powerful tool to investigate the bulk electronic structure~\cite{West1995,Dugdale1997,Dugdale1999,Dugdale2016,Weber2015,Ceeh2016}.
It is based on the annihilation of positrons with electrons in the sample, leading to the emission of two $\gamma$-quanta  in nearly anti-parallel directions. 
The small angular deviation from collinearity is caused by the transverse component of the electron momentum. 
The coincident measurement of the annihilation quanta for many annihilation events yields a projection  of $\rho^{2\gamma}({\bf p})$.
This is usually considered to correspond to the Fourier transform of two-particle electron-positron wave functions. A standard further approximation in the investigation of momentum densities is to factorize the pair wave functions into products of the positron wave function $\Psi^+({\bf x})$ and the electron wave functions $\Psi^-({\bf x})$:
\begin{equation}
	\rho^{2\gamma}({\bf p}) \propto \sum_{j,k} n_j({\bf k}) 
\left| \int \dd{\bf x} \, e^{-i \hbar {\bf x k}} \Psi^+({\bf x}) \Psi_{j,{\bf k}}^-({\bf x}) \, \sqrt{\gamma({\bf x})} \right|^2 .
    \label{eq:rho2gamma}
\end{equation}
The sum runs over all states ${\bf k}$ in all bands $j$ with the occupation $n_j({\bf k})$. The so-called ``enhancement factor'' $\gamma({\bf x})$~\cite{Jarlborg1987}, takes into account electron-positron correlations (typically corresponding to increased contact densities).
A formally equivalent approach in terms of Green's functions is discussed in Sec.~\ref{sec:theor}.
The measured 2D-ACAR spectrum $N(p_x,p_y)$ is a statistical realization of the 2D projection of the 3D momentum-density distribution $\rho^{2\gamma}({\bf p})$ along a chosen ($p_z$) axis convoluted by a 2D angular point spread function $\mathcal{R}(p_x,p_y)$ to account for the spectrometer resolution:
\begin{equation}\label{eq:2dacar}
N(p_x,p_y) = \left[ \int \rho^{2\gamma}({\bf p}) \dd{p_z} \right] \ast \mathcal{R}(p_x,p_y). 
\end{equation}
The main contributions to $\mathcal{R}$ are the position resolution of the detectors and the positron spot size on the sample.
The total resolution (FWHM) of the measurement is \SI{1.45}{\milli\radian} and  \SI{1.12}{\milli\radian} 
in $p_x$- and in $p_y$-direction, respectively.

2D-ACAR spectra were measured using the TUM spectrometer~\cite{Ceeh2013} with a $^{22}$Na positron source.
A single crystalline vanadium rod with \SI{10}{\milli\meter} length, \SI{6}{\milli\meter}
diameter and a purity of 5N was purchased from GoodFellow. It was oriented and cut into a  disc of approximately \SI{1}{\milli\meter} thickness and its surface polished.  With X-ray diffraction we determined the lattice parameter to be \SI{3.028}{\angstrom}. 
Five projections were recorded at room temperature by turning the sample around the $[110]$ axis, collecting ~\num{35e6} coincident counts on average. The projections comprised the main symmetry directions $[001]$, $[1\bar{1}0]$, and $[1\bar{1}1]$ and the projections at \SI{20}{\degree} and \SI{70}{\degree} with respect to the $[001]$ direction.
Parallel to the ACAR measurement the non-coincident \SI{511}{\keV} events were recorded as well to determine the momentum sampling function which accounts for the angle dependent detection efficiency. Distortions in the position assignment of the two-dimensional detectors were corrected by way of a calibration pattern.\cite{Leitner2012}

As the $\rho^{2\gamma}$ has the  symmetry of the cubic crystal, the {ACAR} spectra also possess certain symmetries depending on the projection direction. In particular the $[100]$ projection has the symmetry group of a square $D_4$ (Schoenflis notation). However, due to the anisotropic resolution function, the symmetry is reduced to the two fold symmetry $D_2$. In order to suppress the statistical noise, we took advantage of this fact to obtain a symmetrized spectrum $\tilde{N}(p_x,p_y)$:
\begin{equation}\label{eq:N}
	\tilde{N}(p_x,p_y) = \sum_{g \in D_2} g[N(p_x,p_y)].
\end{equation}

Core states, which the positron probes with reduced weight, lead to a predominantly isotropic contribution to the measured spectra. The \textit{anisotropic} part is primarily due to conduction electrons, and specifically to the Fermi-Dirac occupation function, constituting the aspects of interest to the study reported here. 
This anisotropy  $A(p_x,p_y)$ is brought out more clearly when the isotropic features are subtracted  from the 2D-ACAR spectrum $N(p_x,p_y)$:
\begin{equation}\label{eq:A}
A(p_x,p_y) = \tilde{N}(p_x,p_y) - C(p_x,p_y).
\end{equation}
The radial average  $C(p_x,p_y) \equiv C(\sqrt{p_x^2+p_y^2}) = C(p_r)$  is constructed from the original spectrum $\tilde{N}(p_x,p_y)$ averaging over all data points in equidistant intervals $[p_r,p_r+\Delta p_r)$ from the center.

\section{Theoretical techniques}
\label{sec:theor}

The theoretical analysis of the 2D-ACAR spectra requires the knowledge of the two-particle electron-positron Green's function, describing the probability amplitude for an electron and a positron propagating between two different space-time points.
The theory of the annihilation probability of a positron in a homogeneous electron gas has a long history in many-body physics~\cite{ca.ka.65,carb.67,ar.pa.79,ru.st.88,bo.sz.81}.
The electron-positron attraction leads to an increase of the electron density near the positron. It manifests itself in the annihilation characteristics and leads to a strongly increased total annihilation rate.
This effect is qualitatively well understood and is called ``enhancement''.
However, apart from the short-range screening the electronic states and the mean density remain almost unchanged. 
Therefore, the 2D-ACAR shows only relatively small differences compared to the results of the independent particle model.
In the case of alkali metals the enhancement effect is included by multiplying the 2D-ACAR spectra computed in the independent particle model with an isotropic enhancement factor~\cite{Kahana63,ca.ka.65}, the so-called Kahana factor.
This approach was generalized to an energy dependent form~\cite{Mijnarends79} and was later extended to include orbital dependence~\cite{Sing86}.
Since this was formulated within DFT~\cite{Niem86,ru.st.88,ba.pu.95,ba.pu.96,Barb97,sorman96,Sormann2006,Sznaid12,Sznaid14,Kontrym-Sznajd2014,Boronski2014} the results maintain their static mean-field character.

Electronic structure calculations were performed with the Spin-Polarized Relativistic Korringa-Kohn-Rostoker (SPR-KKR) code~\cite{EKM11}. 
In the LDA computations the exchange-correlation potentials parametrized by Vosko, Wilk and Nusair~\cite{VWN80} were employed. 
The experimental lattice parameter of $3.028\, $\AA, and a BZ-mesh of $22\times22\times22$ was used throughout the calculations. 
The DFT can be generalized to the problem at hand by including the positron density in the form of a 2-component DFT~\cite{Niem86,pu.ni.94}. 
In the present calculations the electron-positron correlations are taken into account by a multiplicative (enhancement) factor [$\sqrt{\gamma({\bf x})}$ in Eq.\eqref{eq:rho2gamma}], which results from the inclusion of the electron-positron interaction in the form of an effective one-particle potential as formulated by Boro\'nski and Nieminen~\cite{Niem86}.

In order to discuss possible correlation effects within the framework of LDA+DMFT~\cite{ko.sa.06,held.07}
the standard methodology is to add to the LDA Hamiltonian ${\mathcal H}_\text{LDA}$ the 
following multi-orbital on-site interaction term:
\begin{equation}
{\mathcal H}_{U} = \frac{1}{2}\sum_{{i \{m, \sigma \} }} U_{mm'm''m'''}
c^{\dag}_{im\sigma}c^{\dag}_{im'\sigma'}c_{im'''\sigma'}c_{im''\sigma}.
\end{equation}
The corresponding many-body problem described by the total Hamiltonian
\begin{equation}\label{eq:H}
{\mathcal H}= {\mathcal H}_\text{LDA} + {\mathcal H}_{U} - {\mathcal H}_\text{DC}  
\end{equation}
is solved using the LDA+DMFT method, where ${\mathcal H}_\text{DC}$ serves to eliminate double counting of the
interactions already included in ${\mathcal H}_\text{LDA}$. Here, $c_{im\sigma}$($c^\dagger_{im\sigma}$) destroys (creates) an electron with spin $\sigma$ on orbital $m$ at the site $i$. The Coulomb matrix elements $U_{mm'm''m'''}$  are expressed in the standard way~\cite{im.fu.98} in terms of three Kanamori parameters $U$, $U'$ and $J$.
In the LDA+$U$~\cite{an.ar.97} scheme a mean-field decoupling of the interaction is employed, whereby electronic correlations beyond the LDA parametrization are eliminated.

To include the electronic correlations we employ a charge and self-energy self-consistent LDA+DMFT scheme which is based on the KKR approach~\cite{mi.ch.05}. Contrary to the Hamiltonian formulation, the KKR implementation of the LDA+DMFT uses the multiple scattering concept: the solution for the single-site problem (single-site, multi-orbital $t$-matrix) includes the local self-energy of the many-body problem, and allows for
the evaluation of the scattering path-operators and the real-space DMFT corrected Green's function needed for the DFT calculation. In the above mentioned implementation, the impurity problem is solved, i.e. the many-body self-energy is constructed,  with a spin-polarized $T$-matrix fluctuation exchange method~\cite{li.ka.97,PKL05}. 
This impurity solver is fully rotationally invariant even in the multi-orbital version and is reliable when the interaction strength is smaller than the bandwidth, a condition which is fulfilled in the case of vanadium. 

Both LDA+$U$ and LDA+DMFT computations require the parametrization of the interaction matrix $U_{mm'm''m'''}$ in terms of the average local Coulomb $U$ and exchange parameter $J$. 
These values of $U$ are sometimes used as fitting parameter.
However, recent developments made it possible, in principle, to compute the dynamical electron-electron interaction matrix elements with a good accuracy \cite{AIG+04}, but with substantial variations associated with the choice of the local orbitals \cite{MA08}. Since the parameter $J$ is not affected by screening it can be calculated directly within the LDA and is approximately the same for all 3d elements, i.e $J$ $\approx$ 0.9\,eV.
We use the value $U$ = 2.3\,eV for the Coulomb parameter which corresponds to the static limit of the screened energy dependent Coulomb interaction computed for vanadium~\cite{AIG+04}, and the Hund exchange-interaction $J$ = 0.9\,eV. 
The LDA+DMFT computations are performed at a temperature of $T=\SI{400}{\kelvin}$, and $n=4096$ Matsubara frequencies $\omega_n= (2n+1) \, \pi \, T$ are included.
The Pad\'e~\cite{vi.se.77} analytical continuation is used to map the self-energies from the Matsubara frequencies onto real energies within the self-consistent KKR-based LDA+DMFT~\cite{ch.vi.03,mi.ch.05}. 
The double-counting correction [see Eq.~\eqref{eq:H}] employed here starts with the LDA electronic structure and replaces the computed self-energy $\mathbf{\Sigma}_{\sigma}(E)$ by $\mathbf{\Sigma}_{\sigma}(E)-\mathbf{\Sigma}_{\sigma}(0)$ in all equations of the LDA+DMFT scheme \cite{li.ka.01}, where the energy $E$ is measured relative to the Fermi energy. 
A detailed description of this commonly used double-counting correction scheme for metals can be found in Ref.~[\onlinecite{pe.ma.03}].

\begin{figure*}
	\centering
	\includegraphics[width=\textwidth]{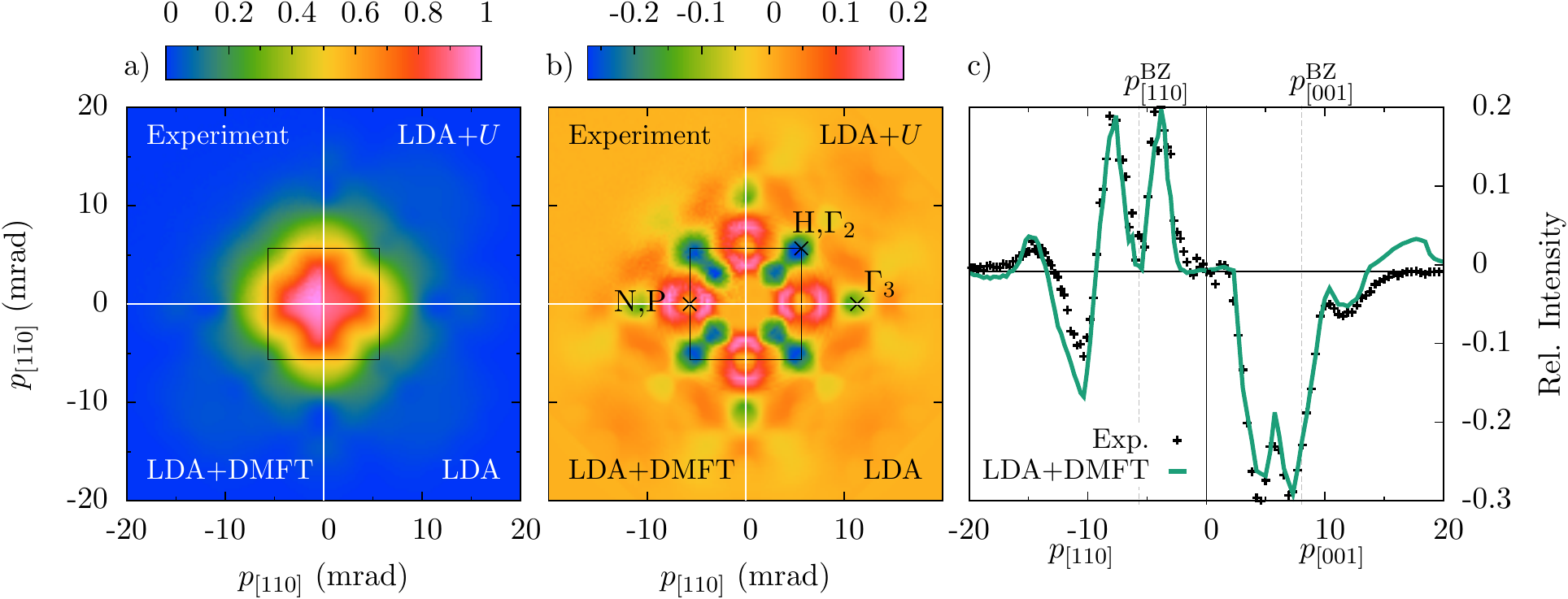}
\caption{Comparison between experimental and theoretical results: (a) 2D-ACAR $\tilde{N}(p_x,p_y)$ spectra; (b) anisotropy $A(p_x,p_y)$ spectra [according to Eq.~\eqref{eq:A}] with integration axis along [100]. The color scale indicates the signal intensity relative to $\tilde{N}(0,0)$. 
The points $\Gamma_2$ and $\Gamma_3$ correspond to the $(1,1,0)$ and $(1,0,0)$ reciprocal lattice vector, respectively.
With respect to the ${\bf k}$-basis the N-point of the Brillouin zone has the coordinates N$(0, \pi/a,\pi/a)$; (c) the cross sections of 2D-ACAR anisotropies ([001] projection) along the [110] and [100] directions illustrate the difference between experimental and DMFT anisotropies. Black crosses: experiment; green curve: LDA+DMFT calculations. The dashed lines indicate the Brillouin zone boundaries  $p^{\text{BZ}}_{[110]/[100]}$ along the [110] and [100] directions, respectively.}
	\label{fig:Vanad:comp}
\end{figure*}

In our LDA+DMFT framework the electron-positron momentum density $\rho^{2\gamma}({\bf p})$ is computed directly from the two-particle Green's function in the momentum representation~\cite{be.ma.06,be.mi.12,ch.be.14}.
The factorization of the pair wave function discussed above in Eq.~\eqref{eq:rho2gamma} is equivalent to the factorization of the two-particle Green's function in real space. 
The enhancement factor $\sqrt{\gamma({\bf x})}$ [see Eq.~\eqref{eq:rho2gamma}] is contained implicitly through the basis of the two-component DFT in which the Green's function is represented. Although this corresponds to the neglect of genuine electron-positron correlations the two-particle Green's function contains correlations between electrons through the DMFT scheme. 
In the numerical implementation the position-space integrals for the ``auxiliary'' Green's function $G_{\sigma \sigma^{\prime}}({\bf p}_\text{e},{\bf p}_\text{p})$ obtained within LDA or LDA+DMFT, respectively, are performed as integrals over unit cells:
\begin{equation}
\label{G_ep}
\begin{split}
G^{X}_{\sigma \sigma^{\prime}}&({\bf p}_\text{e},{\bf p}_\text{p}, E_\text{e}, E_\text{p}) =
\frac{1}{N \Omega}\int \dd[3]{\bf r} \int \dd[3]{{\bf r}^{\prime}} 
\phi_{{\bf p}_\text{e} \sigma}^{\text{e}\dagger}({\bf r}) \times \\ & 
\Im{G^{X}_{\text{e} \  \sigma}({\bf r},{\bf r}^{\prime},E_\text{e})} \,
\phi_{{\bf p}_\text{e} \sigma}^{\text{e}}({\bf r}^{\prime}) \,
\phi_{{\bf p}_\text{p} \sigma^{\prime}}^{\text{p}\dagger}({\bf r}) \times \\ & 
\Im{G_{\text{p}^+ \ \sigma^{\prime}}({\bf r}, {\bf r}^{\prime},E_\text{p})} \,
\phi_{{\bf p}_\text{p} \sigma^{\prime}}^{\text{p}}({\bf r}\,')\;.  \nonumber
\end{split}
\end{equation}
Here $X$ = LDA, LDA+$U$ or LDA+DMFT, and $({\bf p}_\text{e}, \sigma)$, and $({\bf p}_\text{p},\sigma^{\prime})$
are the momenta and spin of electron and positron, respectively. $G^{X}_{\sigma \sigma^{\prime}}$ is computed for each energy point on the complex energy contour, and provides the electron-positron 
momentum density:
\begin{eqnarray} \label{rho_ep}
 \rho_{\sigma}^{2\gamma,X}({\bf p}) &=&  -\frac{1}{\pi} \int \dd{E_\text{e}}
G^{X}_{\sigma \sigma^{\prime}}({\bf p}_\text{e},{\bf p}_\text{p}, E_\text{e}, E_\text{p}). 
\end{eqnarray}
In Eq.~\eqref{rho_ep} integration over positron energies $E_\text{p}$ is not required, since only
the ground state is considered, and $\sigma^\prime = -\sigma$ in the annihilation process. The
momentum carried off by the photons is equal to that of the two particles up to a reciprocal
lattice vector, reflecting the fact that the annihilation takes place in a crystal. Hence an
electron with wave vector ${\bf k}$ contributes to $\rho^{2\gamma, X}_{\sigma}(\textbf{p})$ not only
at ${\bf p} = \hbar\,{\bf k}$ (normal process) but also at ${\bf p} =\hbar \left({\bf k} + {\bf K} \right)$, with
${\bf K}$ a vector of the reciprocal lattice (Umklapp process). From the two-photon momentum density computed for a specific X, $\rho^{2\gamma}(\textbf{p})$, the corresponding 2D-ACAR
spectrum is computed according to Eq.~\eqref{eq:2dacar}.

\section{2D-ACAR anisotropies}

\label{sec:anizo}

The anisotropies, defined as the difference between the 2D-ACAR densities and their radial average [see Eq.~\eqref{eq:A}], arise from the positron-valence electron annihilation, since the extracted radial symmetrized spectra eliminate the core contributions.
Fig.~\ref{fig:Vanad:comp} presents the 2D-ACAR and the anisotropy spectra together with the results of the LDA, LDA+$U$ and LDA+DMFT calculation. 
Our analysis focuses on the two-dimensional projections of the electronic momentum densities with the integration direction chosen along $[100]$. 
The general features of the experiment, such as peaks along the studied directions, are well reproduced by the calculations. 
These peaks originate from the occupation of the single-electron states.
The measured anisotropies are slightly weaker than the calculated ones. 
This is explicitly seen by taking cross sections of 2D-ACAR anisotropies as shown in Fig.~\ref{fig:Vanad:comp} (c).
The calculations overestimate the anisotropy peaks. However, the difference between experiment and theory is rather small, and within the LDA+DMFT it amounts to less than $5\%$ along the [110] and less than  $3\%$ along [100].

The LCW procedure~\cite{Lock1973} can be used to study the Fermi surface in the momentum densities by folding the data into a single central Brillouin zone.
This technique enhances discontinuities by superposing Umklapp terms. 

\begin{figure}[t]
	\centering
    \includegraphics[width=\linewidth]{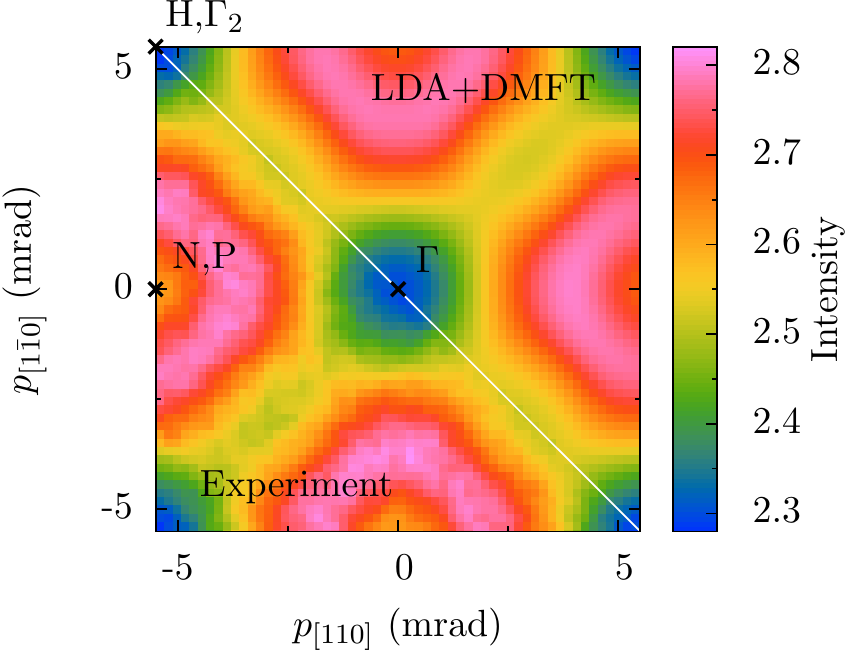}
    \caption{Results for the LCW folding of the 2D-ACAR spectra. Below diagonal: experimental results.
    Above diagonal: LDA+DMFT results. The theoretical spectra was convoluted with the experimental resolution according to Eq.~\eqref{eq:2dacar}.}\label{fig:backfold}
\end{figure}

In Fig.~\ref{fig:backfold} we compare the 2D-ACAR folded experimental spectra with the corresponding LDA+DMFT calculation up to the zone boundary of the first Brillouin zone i.e.  $p^{\text{BZ}}_{[110]}=\SI{5.67}{\milli\radian}$. 
The agreement between the LDA+DMFT results (convoluted with the experimental resolution) and the experimental data is remarkable and is far better than in the case of LDA or LDA+$U$.

\begin{figure}[h]
	\centering
	\includegraphics[width=0.99\linewidth]{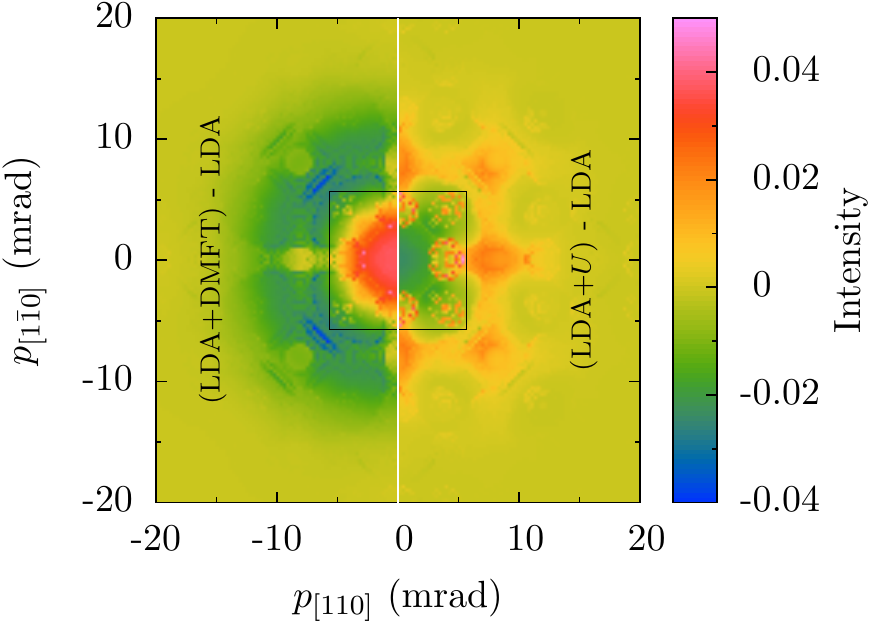}
\caption{Corrections to the theoretical 2D-ACAR spectra ([001] projection) defined as the difference between the LDA+DMFT and LDA calculations (left panel) and the LDA+$U$ and LDA calculations (right panel) respectively.}
\label{fig:corection}
\end{figure}

We investigated the following two possibilities to explain the discrepancies between the LDA/LDA+$U$ results and the measured anisotropies. 
Taking into account the temperature simply by replacing $n({\bf k})$ of Eq.~\eqref{eq:rho2gamma} with the Fermi function leads to no significant change. 
Another possible reason for the discrepancies is the form of the exchange-correlation potential used in the DFT calculations. 
Therefore we computed the corrections to the 2D-ACAR due to the presence of a Hubbard $U$ term in the DFT calculations, by using the LDA+$U$ and LDA+DMFT methods. 
For a quantitative analysis we present the difference spectra $N_{X}(p_x,p_y) - N_\text{LDA}(p_x,p_y)$ in Fig.~\ref{fig:corection}, for the $X$ = LDA+DMFT and LDA+$U$.

As can be seen in Fig.~\ref{fig:corection} there is a significant difference between the results obtained for a static correction (LDA+$U$)  compared to dynamic correlations $\Sigma(i\omega_n)$ in DMFT. For the following qualitative discussion we consider only the orbital diagonal part; non-diagonal parts are omitted to simplify the equations.
The real and imaginary part of the electronic self-energy contribute to the single-particle spectral function as
\begin{equation}\label{eq:Akomega}
A({\bf k}, E) \approx \frac{Z_{\bf k}}{\pi} \frac{\tau_{\bf k}}{(E -\xi_{{\bf k}})^2 + (\tau_{\bf k})^2} +  A_{\text{incoh}}({\bf k}, E).
\end{equation}

Here $1/\tau_{\bf k} = \Im{\Sigma^\text{DMFT}(E)}$ is the scattering rate, and $\xi_{{\bf k}}= \epsilon^\text{LDA}({\bf k}) - \Re{\Sigma^\text{DMFT}(E)}$ is the renormalized energy of the quasiparticle.  The second term in Eq.~\eqref{eq:Akomega} represents the continuous, incoherent part of the spectra, which has to be present in view of the renormalization factor $Z_{\bf k} \le 1$, and for the spectral function sum rule to be satisfied. Accordingly, the frequency integral of the spectral function is the occupation probability of the single-particle state: 
$n({\bf k})=\int_{-\infty}^{E_\text{F}} A({\bf k}, E) \dd{E}$. Fermionic systems in which this picture holds are called Fermi liquids. In a Fermi liquid $1/\tau_{\bf k}$ approaches zero as ${\bf k}$ approaches ${\bf k}_\text{F}$. 
In fact, the imaginary part of the self-energy for real energies (not shown) follows a quadratic energy dependence in the vicinity of the Fermi level: $\Im{\Sigma^\text{DMFT}(E)} \propto (E-E_\text{F})^2$. 
As a consequence of the negative slope of the real part of the self-energy, the quasiparticle energy $\xi_{{\bf k}}$ is shifted towards the Fermi level from its LDA eigenenergy $\epsilon^\text{LDA}({\bf k})$.
For this reason the self-energy correction to the 2D-ACAR spectra leads to a concentration of the momentum density around the $\Gamma$-point (Fig.~\ref{fig:corection}, left panel).
Our results are in accord with Fermi liquid theory and imply that fermions close to the Fermi surface scatter very little.

The effects produced by the orbital dependent potential of the LDA+$U$ method may be viewed as a static self-energy correction: $\Sigma^{\text{LDA}+U}_{m,\sigma} = U_{ml} n_{l,-\sigma}$; in the following we consider only the orbital diagonal case. In the present non-magnetic case the self-energy is spin independent and amounts to a constant shift of the LDA eigen-energy $\epsilon^{\text{LDA}}({\bf k}) - \Sigma^{\text{LDA}+U}_{m}$ for a specific orbital $m$  of the ($d$-)manifold, implying an orbital dependent potential $ \Sigma^{\text{LDA}+U}_{m} = \frac{1}{2}U_{ml} n_{l}$.  Consequently, the momentum density is depleted in the central part of the Brillouin zone and shifted into the second and further Brillouin zones (Fig.~\ref{fig:corection}, right panel). 

\begin{figure}[t]
	\centering
    \includegraphics{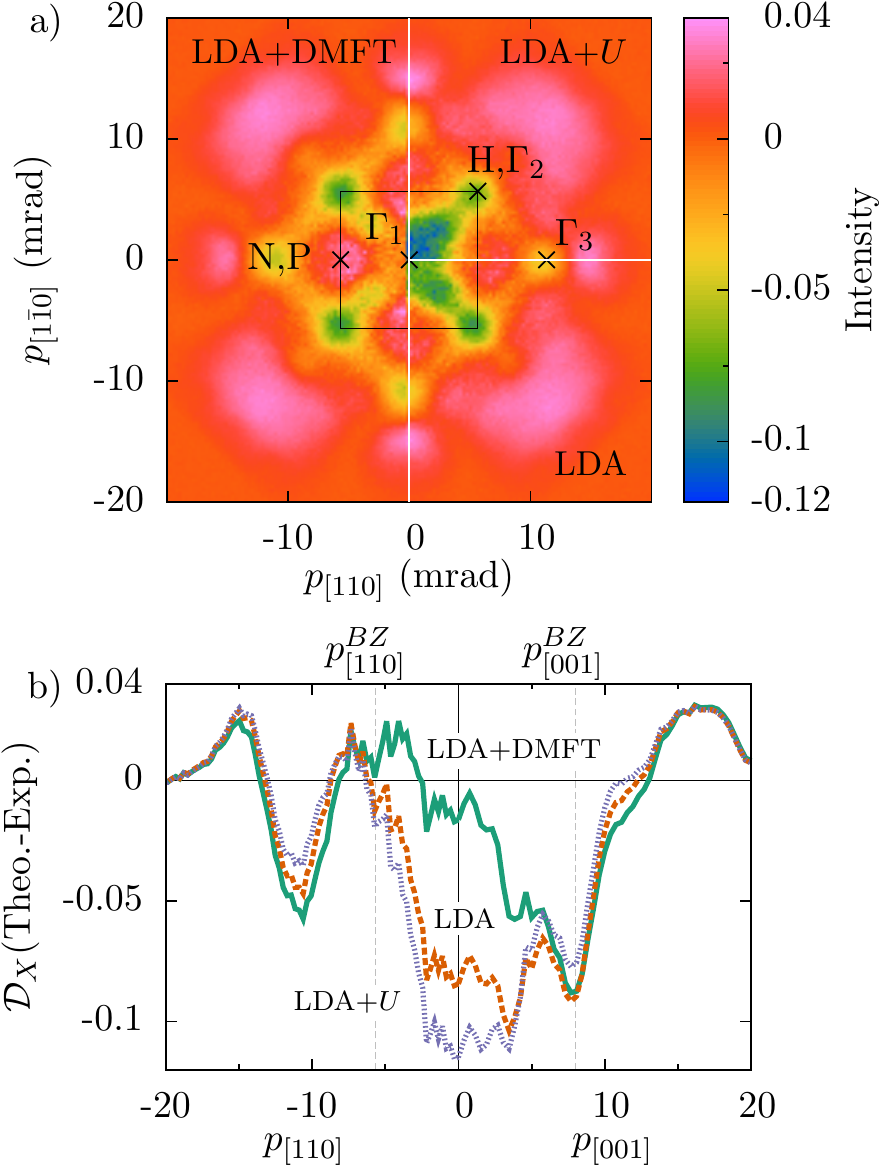}
    \caption{(a) Difference between spectra in theory and experiment $\mathcal{D}_X(\text{Theory}-\text{Experiment})$, see Eq.~\eqref{Delta}. The projections of the  high symmetry points  N $[0,\pi/a,\pi/a]$, 
 P $[\pi/a, \pi/a, \pi/a]$,  H $[0,0,2\pi/a]$, $\Gamma_{1,2,3}$ are marked in the figure.   
The subscript index of the $\Gamma_i$-point, $i=1,2,3$ indicates that these points belong to the first, second and third Brillouin zone. For the $[001]$ projection, the points $N$ and $P$, as well as $\Gamma_2$ and H coincide. (b) Cross sections through the upper picture along $[100]$ and $[110]$; green solid line: LDA+DMFT, orange dashed: LDA, blue fine dashed: LDA+$U$.}\label{fig:exp-theor}
\end{figure}

Fig.~\ref{fig:exp-theor} shows the results for the difference of the calculated and the experimental spectra 
$\mathcal{D}_X$:
\begin{equation}\label{Delta}
{\mathcal{D}}_X=N_{X}(p_x,p_y) - N_\text{exp}(p_x,p_y)
\end{equation}
On the left hand side of Fig.~\ref{fig:exp-theor}(a), the 
difference spectra ${\mathcal{D}}_\text{LDA+DMFT}$ is shown. Except for the H-point
the weight of the theoretical spectra (intensity) differs from the experimental data only slightly (maximum difference intensity $\pm 0.04$). On the right hand side of Fig.~\ref{fig:exp-theor}(a), the differences computed with LDA and LDA+$U$ results are presented. Along the $\Delta$ (i.e. $\langle 1 0 0 \rangle$) direction, connecting the centers of the first and the second Brillouin zone ($\Gamma_1( \equiv \Gamma)$ and $\Gamma_2$), both LDA and LDA+$U$ show more significant departures from the experimental data. This hole-like feature is part of the open Fermi surface sheet referred to as ``jungle-gym''~\cite{Mattheiss1970}, shown also in  the reconstructed 3D Fermi surface, Fig.~\ref{fig:vanadiumReconstruction}. 
In the spectra it is resolved as a low intensity region. 
Concerning the size of the jungle-gym obtained from different functionals we refer to Fig.~\ref{fig:spektral}. Our results show that going along the $\Gamma$H-branch the LDA+$U$ jungle-gym is  narrower than in the LDA calculation.

The jungle-gym can also be seen very clearly along the $\Delta$ direction in the anisotropy [Fig.~\ref{fig:Vanad:comp}(b)]. A second distinct detail is the low intensity pocket at the projected (N,P)-point\footnote{In the [100]-projection N and P are projected to the same point. In Fig.~\ref{fig:vanadiumReconstruction} a different section is presented in which the N and P points are distinct.}. 
It originates from the ellipsoidal hole pocket at the N point, seen also in Fig.~\ref{fig:vanadiumReconstruction}.  
The same feature is also present in the {LCW} back-folded spectra, see Fig.~\ref{fig:backfold}. 
We observe in Fig.~\ref{fig:exp-theor} that the differences ${\mathcal{D}}_\text{LDA+DMFT}$ and the ${\mathcal{D}}_\text{LDA}$  agree with experiment quite well at this particular high symmetry point of the Brillouin zone. Again the LDA+$U$ result ${\mathcal{D}}_{\text{LDA}+U}$ overestimates the N-pocket. 
Therefore the pocket is larger than in the experiment. 
The same conclusion can be drawn form the 3D reconstructed Fermi sheets discussed in Sec.~\ref{sec:fs}.

In the lower part of Fig.~\ref{fig:exp-theor} cross sections of the difference spectra ${\mathcal{D}}_{X}$ along $\langle 1 0 0 \rangle$ and $\langle 1 1 0 \rangle$ are presented. 
The vertical dashed lines indicate boundaries of the first Brillouin zone. From these plots it becomes obvious
that most differences arise within the first Brillouin zone. The best agreement between the LDA+DMFT calculations and the measurements is found around the $\Gamma$ point. The agreement becomes less good towards the zone boundary, where the results coincide with those from LDA and LDA+$U$ in both directions along $\langle 0 0 1 \rangle$ and also along $\langle 1 1 0 \rangle$.

\section{Fermi surface results}

\label{sec:fs}

In spite of intensive studies, both theoretically and experimentally, only very limited information about the Fermi surface of vanadium is available.
Although reconstructions of the Fermi surface were previously reported\cite{Manuel1982,Major2004a} they provide only a qualitative comparison between experimental and calculated results. Two exceptions are the {dHvA} measurements of Phillips \cite{Phillips1971} and the theoretical study of Tokii and Wakoh \cite{Tokii2003}. 
Phillips~\cite{Phillips1971} was able to resolve the shape of the ellipsoid N-hole pockets and gave the dimensions of the semi-axis NP, N$\Gamma$ and NH as 0.224, 0.212 and 0.176 in units of the reciprocal lattice constant\cite{Phillips1971}. Tokii and Wakoh~\cite{Tokii2003} performed {LDA} and {LDA}$+U$ calculations with which they could reproduce the results of Phillips \cite{Tokii2003}. 

\begin{figure}[h]
	\centering
	\includegraphics[width=0.99\linewidth]{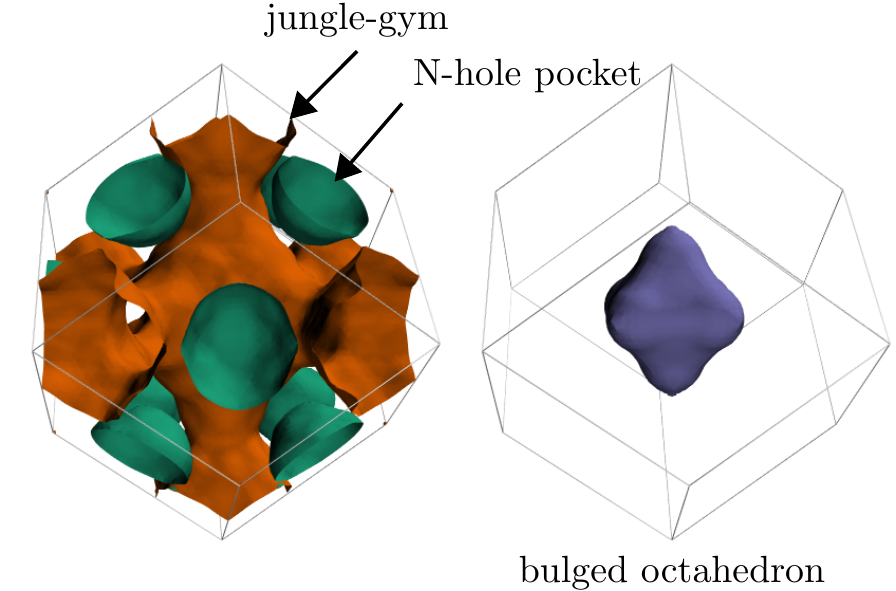}
	\caption{Reconstructed Fermi surface of vanadium. All sheets are hole-like, i.\,e. the occupied states are between the ellipsoidal N-hole pockets (green) and the jungle-gym (orange), and outside of the octahedron (violet). Note that the jungle-gym and the N-hole pocket are drawn with the same iso-value but were depicted with different colors to obtain a better visibility.}
	\label{fig:vanadiumReconstruction}
\end{figure}

Using the five measured projections, we here recover the full 3D $\rho^{2\gamma}$ by an iterative algebraic reconstruction technique~\cite{Weber2015} which takes into account the full symmetry of the Brillouin zone and the anisotropic resolution of the spectrometer. The reconstruction of ${\rho}^{2\gamma}({\bf p})$ is computed by minimizing the squared difference between the measured projections $N_i(p_x,p_y)$ and the projections of a test density ${\rho}^{2\gamma}({\bf p})$
\begin{equation}
{\rho}^{2\gamma}=\argmin_{{\rho}^{2\gamma}} \sum_i
\left[\frac{\left( N_i- \sum_j P_{ij} {\rho}^{2\gamma}_j
\right)^2}{\bm{\sigma}_i^2} +
{\rho}^{2\gamma}_i\ln{{\rho}^{2\gamma}_i}\right].
\end{equation}
The projection operator $P_{ij}$ comprises a projection from the irreducible wedge, a convolution with the experimental resolution, the detection efficiency in momentum space, and a scaling; $\sigma_i$ denotes the expected uncertainty. The entropy-like regularization is needed since the problem is underdetermined.
Using the {LCW} procedure for 3D, $\rho^{2\gamma}({\bf p})$ was then folded back into ${\bf k}$-space to obtain $\rho^{2\gamma}({\bf k})$. 
As the LCW theorem does not hold exactly due to positron wave function and correlation effects,  $\rho^{2\gamma}({\bf k})$ is not flat but shows smooth variations. O'Brien et al.~\cite{OBrien1995} accounted for this variation by a bandpass filter. In this study, we chose a different approach: We modeled the variations by the first three Fourier coefficients of the lattice. In this way we were able to enhance the Fermi breaks. Finally, {iso-values} were chosen at the minima of the histogram of $\rho^{2\gamma}({\bf k})$ in order to draw the Fermi surface. The reconstructed FS sheets of vanadium are shown in  Fig.~\ref{fig:vanadiumReconstruction}.

The reconstruction reproduces all features of the vanadium Fermi surface discussed in earlier papers~\cite{Major2004a,Singh1985a,Dugdale2013}. In the center of the Brillouin zone a hole-like volume is confined by the bulged octahedron FS sheet (violet in Fig.~\ref{fig:vanadiumReconstruction}).  
Further small hole-like FS sheets, the so-called ``N-hole pockets'', are centered at every N-point of the Brillouin zone (green in Fig.~\ref{fig:vanadiumReconstruction}). The largest {FS} sheet  has tube like arms which connect with neighboring Brillouin zones at the X-points (orange in Fig.~\ref{fig:vanadiumReconstruction}).  When drawn in the repeated zone scheme this sheet forms a regular grid and was therefore termed ``jungle-gym''~\cite{Mattheiss1970}. Both the jungle-gym an the N-hole pocket give very pronounced features in the $[001]$ projection (for a detailed discussion see  Sec.~\ref{sec:anizo}).

\begin{figure}[h]
	\centering
    \includegraphics[width=0.55\linewidth]{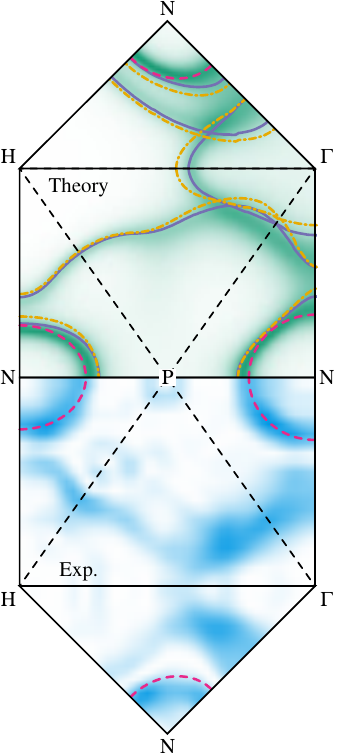}
    \caption{Upper half: LDA+DMFT spectral function of vanadium together with the LDA (solid violet) and LDA+$U$ (yellow dot-dashed curve) Fermi surfaces. Lower Half: Experimental data, representing the absolute squared gradient of the reconstructed and LCW folded $\rho^{2\gamma}$ according to Eq.~\eqref{gradrho}.
The {dHvA} results by Phillips~\cite{Phillips1971} are shown with purple dashed line.}
    \label{fig:spektral}
\end{figure}

Additionally, we plot in Fig.~\ref{fig:spektral} the absolute square of the gradient of the reconstructed function $\rho^{2\gamma}({\bf k})$:
\begin{equation}\label{gradrho}
|\nabla \rho^{2\gamma}({\bf k})|^2 = \left( \frac{\partial \rho^{2\gamma}}{\partial {\bf k}_x} \right)^2 + \left( \frac{\partial \rho^{2\gamma}}{\partial {\bf k}_y} \right)^2 + \left( \frac{\partial \rho^{2\gamma}}{\partial {\bf k}_z} \right)^2 .
\end{equation}

As the Fermi surface causes discontinuities in $\rho^{2\gamma}({\bf k})$, a high gradient indicates the presence of a Fermi break. This visualization has a technical advantage compared to the drawing of isolines: Because of the finite resolution of the spectrometer, isolines cannot cross but rather repel each other. By contrast, the absolute square of the gradient can capture this band crossings quite well.

In Fig.~\ref{fig:spektral} planes on the boundary of the irreducible element of vanadium's  Brillouin zone are shown.
The triangular section $\Gamma$HN corresponds to the central $\{100\}$ plane, and the rectangular section $\Gamma$HNPN corresponds to the  $\{110\}$ planes. The region around the $\Gamma$ point corresponds to the section through the octahedron shown in Fig.~\ref{fig:vanadiumReconstruction}. 
The octahedron around the $\Gamma$ point is contained within the jungle-gym surface, whose branch extends along H$\Gamma$. The Fermi surface section through the ellipsoids at the N-point is also visible. 
In Fig.~\ref{fig:spektral} the green broadened contours represent the LDA+DMFT results for the spectral function; their width is proportional to the imaginary part of the self-energy. The Fermi surface results obtained by LDA and LDA+$U$ are represented by dashed and dot-dashed curves.  
According to the LDA and LDA+$U$ calculations, the jungle-gym and the octahedral Fermi-sheets touch at three points: one contact point is located in the $\{100\}$ plane and two others in the $\{110\}$ - plane. Along $\Gamma$P in the $\{110\}$ plane one of the contact points corresponds to a degeneracy induced by symmetry, while the other two contact points are caused by accidental degeneracies of states of even and odd reflection symmetry in these planes. 
 
As can be seen, our experimental results are in very good agreement with the elliptical parametrization of the N-hole pocket by Phillips~\cite{Phillips1971}. It should be noted, that his parametrization necessarily results in a simplification of the true shape of the N-hole pocket. Hence, it is not surprising that our results deviate in some details from an ellipsoid.

\section{Conclusion}
\label{sec:sum}
We performed 2D-ACAR measurements of the two-photon momentum density in order to determine the full 3D Fermi surface of a high quality single-crystal of vanadium. 
The reconstructed Fermi surface comprises three {FS} sheets which contain closed hole pockets centered at the symmetry points $\Gamma$ and N of the bcc Brillouin zone as well as a multiply connected hole-sheet extending from $\Gamma$ to H in the $\langle 1 0 0\rangle$ direction.

In order to provide a theoretical model for the Fermi surface of vanadium we performed LDA+DMFT, LDA and LDA+$U$ calculations.
In particular, the agreement between the DMFT results and the experimental data is remarkable and significantly better than in the case of LDA or LDA+$U$.
We observed that LDA and LDA+$U$ underestimate the size of the jungle-gym while it overestimates the N-pocket compared to our LDA+DMFT calculation.

Compared to the measurements of Phillips~\cite{Phillips1971} the N-hole pocket in our reconstruction (Fig.~\ref{fig:spektral}) is slightly shortened in NP direction, while the other semi-axes of the ellipsoid agree very well. 
In contrast to the reconstruction of Manuel \cite{Manuel1982}, the N-hole pocket agrees much better with the theoretical prediction. 
Several details of our calculation are reproduced by the reconstruction. 
The jungle-gym \eg widens from $\Gamma$ to H and has a larger diameter in the HN direction than in the HP direction. 
Hence, with the 3D reconstruction we succeeded to confirm several details predicted by theory.

The quantitative agreement between the calculation and our experimental 2D-ACAR spectra and the anisotropy is excellent in the case of LDA+DMFT, indicating that vanadium is a conventional, correlated Fermi-liquid. 
It is interesting to note that in the case of vanadium the LDA+$U$ does not improve the agreement with the experiment but gives a worse result than the pure LDA scheme. 

Our results indicate that electron-electron correlations have a significant effect. 
Although we treat the direct electron-positron interaction within the usual LDA based approach, we see that the $\rho^{2\gamma}$ contains correlation effects, predominantly of electronic origin, as they come through the self-consistent charge and self-energy two-component DFT calculation. 
Therefore, the consequences of electron-positron correlation in the ACAR distribution need to be investigated in more detail. 
The remaining discrepancies, may be due to the positron self-energy effect~\cite{fu.hy.72}, or the positron dependent enhancement factor.

\section*{Acknowledgements}
This project is funded by the Deutsche Forschungsgemeinschaft (DFG) within the Transregional
Collaborative Research Center TRR 80 ``From electronic correlations to functionality''.
D.B. acknowledges financial support provided through UEFISCDI grant
PN-II-RU-TE-2014-4-0009 (HEUSPIN). We would like to thank H. Ebert and J. Min\'ar
for fruitful collaboration especially for providing the KKR+DMFT package.

\bibliography{sample}
\end{document}